# Ferro-deformation and shape coexistence over the nuclear chart: 28 ≤ protons (Z) ≤ 50 and 40 ≤ neutrons (N) ≤ 70

Chang-Bum Moon*

*Hoseo University, Chung-Nam 336-795, Korea*

(April 11, 2016)

With the experimental data at the national nuclear data center, NNDC, we investigate systematically the emerging nuclear structure properties in the first $2^+$ excited energies, $E(2^+)$ and their energy ratios to the first $4^+$ levels, $R = E(4^+)/E(2^+)$ in the nuclei over $28 \leq Z \leq 50$ for protons, and $40 \leq N \leq 70$ for neutrons. By introducing the *pseudo-shell* configurations built on the combined subshells, we explain the following phenomena; a semi-double shell closure, a shape phase transition, and a reinforced deformation. The reinforced deformation arises suddenly at Z = 40 (or 38), N = 60 and approaches a maximum value, R = 3.3, as being centered at Z = 40, N = 64. We define this reinforced deformation 'a ***ferro-deformation***', as in the previous study [arXiv:1604.01017]. The shape coexistence would be expected to be, as forming a ferro-deformation, with a strong rotational mode, and a near spherical shape, with a vibrational mode, in the transitional region at N = 58, 60, and 62 for the nuclei, with Z = 38 and 40; $^{96}$Sr, $^{98}$Sr, $^{100}$Sr, and $^{98}$Zr, $^{100}$Zr, and $^{102}$Zr. We suggest that the ferro-deformation should be closely associated with a strong spin-orbital interaction between neutrons in the $\mathbf{g_{7/2}}$ orbital and protons in the $\mathbf{g_{9/2}}$ orbital. Such an isospin dependent spin-orbital interaction, with the same angular momentum, $l = 4$, reinforces nuclear surface toward a sudden and dramatic deformation, giving rise to the ferro-deformation at the critical point, Z = 40, N = 64. We discuss the similarities and differences of the ferro-deformation between the two critical points; Z = 40, N = 64 and Z = 64, N = 104. We draw a conclusion that proton-neutron interactions in the spin-orbit doublet play an essential role in controlling nuclear shape phase transitions. Moreover, this interaction should be assigned a special role to constraints on nucleosynthesis in the rapid neutron capture processes.

**Keywords**: national nuclear data center (NNDC), nuclear shell closure, pseudo-(sub)shell, shape phase transition, ferro-deformation, shape coexistence, isospin dependent spin-orbital interaction.
**Nuclides**: Ni, Zn, Ge, Se, Kr, Sr, Zr, Mo, Ru, Pd, Cd, Sn.

*cbmoon@hoseo.edu

## 1. Introduction

Atomic nuclei are complex quantum-mechanical many-body systems with a finite number of nucleons, protons, Z, and neutrons, N. The finite size of a nucleus, which is a mesoscopic system comparing to a microscopic few-body system and a macroscopic system such as a condensed matter, plays an essential role in controlling their properties. In a sense, the shell structure of the atomic nucleus is one of the cornerstones for a comprehensive understanding of the many-body quantum systems. To observe the behavior related to such mesoscopic quantum systems, it is instructive to draw systematically experimental observables according to the number of nucleons [1-6]. Especially, systematic studies of the long isotones and isotopes sequences of the even-even nuclei across the major shell closures are crucial for understanding the nuclear shell structure evolution. Figure 1, together with the shell model level schemes, shows the systematics of the first $2^+$ excited level energies, $E(2^+)$, over the nuclides, Z = 28 to 50 and N = 40 to 70. The distinctive shell structure features are clearly revealed such as a large shell gap at N = 50.

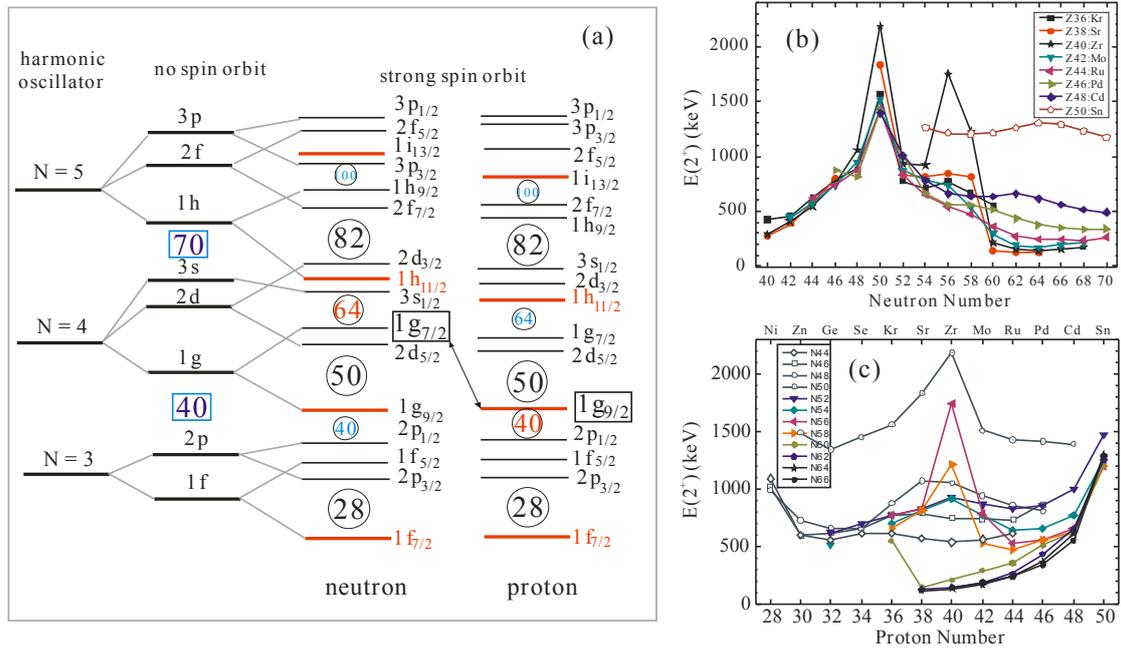

Fig. 1. (a) The single particle energies of a harmonic oscillator potential as a function of the oscillator quantum number N; a schematic representation of the single-particle energies of a Woods-Saxon potential; and a schematic illustration of the level splitting due to the spin-orbit coupling term. The numbers at the energy gaps are the subtotals of the number of particles represented by $N_j = 2j + 1$ of identical particles that can occupy each state. Note that the levels are numbered serially by a given orbital quantum number. For discussions, we denote the associated spin-orbit doublet by the arrow and the corresponding critical points of Z = 40 and N = 64 are in red color. (b) Systematics for the first $2^+$ excited states in the nuclides over Z = 38 to 50 as a function of neutron numbers and (c) over N = 44 to 66 as a function of proton numbers. Data are taken from NNDC [7].

With the aid of the nuclear data at the national nuclear data center (NNDC) [7], we examine the characteristics of nuclear level structures on the basis of experimental observables in the even-even nuclei. By extracting systematic behavior of the first $2^+$ level energies, $E(2^+)$, and the ratios of the first $4^+$ level and $2^+$ level energies, R = E(4+)/E(2+), we discuss the various nuclear structural phenomena over the nuclear chart, 28 ≤ Z ≤ 50 and 40 ≤ N ≤ 70: pseudo-shell configurations, ferro(reinforced)-deformation, shape phase transitions, and shape coexistence. The concepts of ferro-deformation will be explained in terms of isospin dependent spin-orbital interactions between neutrons in the $g_{7/2}$ orbital and protons in the $g_{9/2}$ orbital. This neutron-proton spin-orbit doublet level scheme provides us with an effective visualization of the pathways, as shown in Fig. 1(a), indicating nuclear shape phase transitions at a particular critical point, for instance, Z = 40, N = 64.

## 2. Ferro-deformation and shape coexistence

We begin by describing the systematic behavior of low-lying level properties in the even-even nuclei, as already mentioned, in terms of the excitation energy ratios of the first $2^+$ and $4^+$ states, R = $E(4^+)/E(2^+)$. This value, as a deformation parameter, evolves from < 2 for a spherical nucleus through 2 for a vibrator, to 3.3 for a deformed axial rotor [6, 8]. Figure 2 illustrates the systematics of the R values for a given nuclide, along N = 40 to 70. For comparison, the $2^+$ energy systematics is also included.

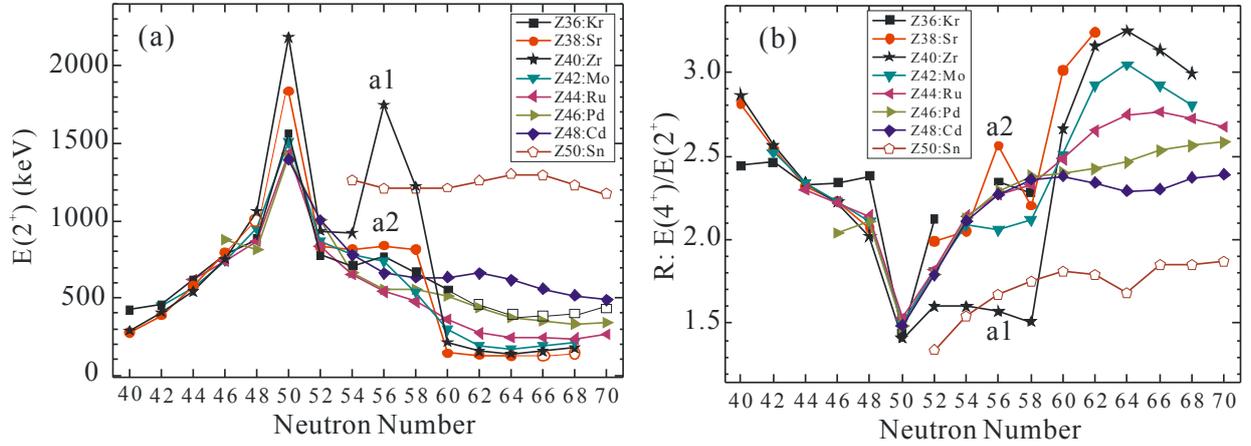

Fig. 2. (a) Systematic plots of $E(2^+)$ and (b) R, $E(4^+)/E(2^+)$, values as a function of neutron numbers for the nuclei within Z = 36 and 50.

In these R and $E(2^+)$ systematics, we notice firstly a sudden transition between N = 58 and 60 for Z = 38 and 40. In turn, at N = 62 the deformation is more enhanced and finally is maximized at N = 64. We remember that the nucleon number 64 corresponds to a semi-shell gap, giving the pseudo-subshell in our previous work [9]. At N = 64 we can see also shape transitions from Z = 48 to Z = 44, through a typical transitional character at Z = 46. Especially the change of Z = 40, Zr, is dramatic; from R = 1.5, a spherical shape at N = 58, through 2.7, a moderate deformation at N = 60, finally R = 3.2, a strong deformation at N =62. Such a sudden and dramatic changes are also seen in $E(2^+)$ systematics. In view of characteristic of the $E(2^+)$ parameter for the case of Z = 40, three shell gaps are formed at 50, 56, and 58, respectively. In fact, the system with Z, N = 40, 50 shows a double-shell closure, very similar to that of Z, N = 20, 28; $^{48}$Ca. It should be remarked further with respect to Z = 40 at N = 56, as being denoted a1 in Fig. 2, or at N = 58 as though it is weaker than at N = 56, the $2^+$ level energy and R values characterize a semi-double shell closure. In contrast, such a semi-shell gap feature is not seen in Z = 38 and 42, as coming out a collective character at N = 56, a2 in Fig. 2. The comparable R values for Z = 40 at N = 52 and 54 to those at N = 56 and 58 attract our attention as well since the $E(2^+)$ values are relatively so low.

Now we turn our attention to the systematics for isotones, as shown in Fig. 3, as a function of the number of protons. The emerging characteristics are as follows: First, for isotones within the N = 50 and 58 space, a strong shell gap occurs at Z = 40, which generates two *pseudo subshells*. One pseudo-shell with Z = 28 to 40, as denoted $J_{pf}(11/2)$, has the capacitance of 12 nucleons occupancy and the other, within Z = 40 and 50, $J_g(9/2)$, has the 10 nucleons. Below N = 50, in contrast, such a shell gap at Z = 40 is not seen. Surprisingly, the two pseudo-subshells suddenly disappear at N = 60, revealing one shell character with a peak at Z = 38 or 40. This one shell is the pseudo shell, as denoting $J_{pfg}(21/2)$, with a capacitance of 22 nucleons. It is worth knowing that the half-occupancy for this pseudo-shell corresponds to Z = 38.

An abrupt change between N = 58 and 60, as already shown in Fig. 2, offers a clear signature for a shape phase transition at and near Z = 40. This Z = 40 and N = 60 number over the nuclear chart corresponds to a critical point occurring one shell structure for both protons and neutrons; $J_{pfg}(21/2)$ within Z = 28 and 50 for protons, and $J_{dghs}(31/2)$ within N = 50 and 82 for neutrons. Provided that half-occupancy at Z = 38 and N = 64, both $J_{pfg}(21/2)$ and $J_{dghs}(31/2)$ pseudo-shells drive a strong deformation, leading to R = 3.3. As shown in Fig. 3, this explains why the Sr (Z = 38) has a maximum value at N = 64. This reinforced deformation has been commonly identified in the region of Z = 64 and N = 104 and we defined it a ferro-deformation [9]. Similarly, we propose that the ferro-deformation should be assigned to the systems; Z = 38, 40, N = 64, 66. Compared with phase transitions for Z = 64 (Gd) at N = 88 and N = 90 [9], the feature of Zr at N = 58 to 60 shows a more dramatic change, suggesting that the Z = 40 subshell closure is stronger than at Z = 64. See the next figure, Fig. 5. From the present point of view, we expect that no semi-double shell closure would appear to be at Z = 40 and N = 70, $^{110}$Zr. The reason is that the nucleon number 70, known as a shell gap number based on a harmonic oscillator potential, turns out not to contribute to producing any semi-subshell.

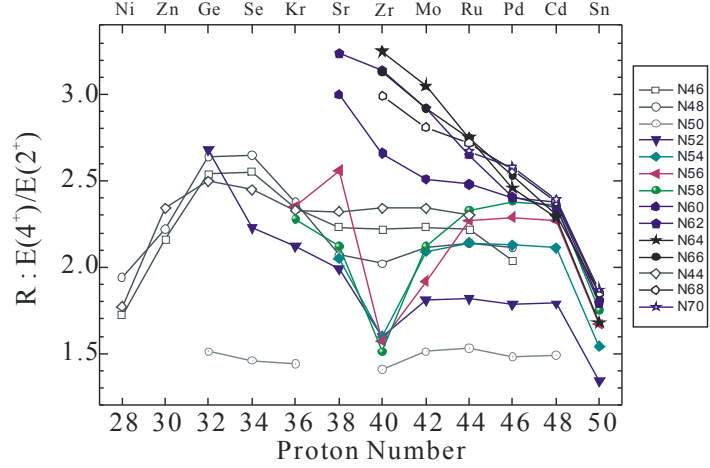

Fig. 3. Systematic plots for R (= E(4$^+$)/E(2$^+$)) values as a function of proton numbers for isotopes between N = 44 and 70.

We suggest that shape coexistence would occur, such as a ferro-deformation, with a strong rotational mode, and a near spherical shape, with a vibrational mode, at the critical points of Z = 38, 40, N = 58, 60, 62; $^{96}$Sr, $^{98}$Sr, and $^{100}$Sr and $^{98}$Zr, $^{100}$Zr, and $^{102}$Zr. Especially, the $^{98}$Sr and $^{100}$Zr are expected to have triple coexistence; a spherical shape, R ~ 1. 6 for Zr and 2.0 for Sr, a moderate deformed shape, R ~ 2.6, and a highly deformed shape, a ferro-defromation, R ~ 3.2. We argue that this ferro-deformation could correspond to a super-deformation, which can be formed at high-lying excited states, by confirming shape coexistence for Sr and Zr, with N = 58 or 60.

We notice additionally that the Cd nuclei, Z = 48, with N < 60 indicate their deformation parameters to be similar to those at Z = 44 and 46. Taking a strong shell gap effect due to Z = 50 into consideration, on one hand, this aspect seems to be extraordinary. On the other hand, this anomaly may be explained by a weak deformation at Z = 46 and 44. This controversy is likely related to the interplay between hole-like protons in the g$_{9/2}$ orbital below Z = 50 and neutrons in the g$_{7/2}$ orbital above N = 50, leading to different impacts on deformation.

We raise a question why the ferro-deformation arises suddenly at the critical points. An answer can be made by introducing the spin-orbital interactions between protons and neutrons in the spin-orbital doublet with the same orbital angular momentum, $l$ = 4; the g$_{7/2}$ for neutrons and the g$_{9/2}$ for protons. This isospin dependent spin-orbital interaction between protons Z < 50 and neutrons N > 50 induces a sudden and dramatic shell structure change when N = 58 to 60. This interaction eventually gives rise to the ferro-deformation at the critical points; Z = 38, N = 64 or Z = 40, N = 64, by forming a huge neutron pseudo-shell, J$_{dghs}$(31/2), which is simultaneously reinforced with the half-filled protons in the pseudo-shell J$_{pfg}$(21/2). We remember that the critical points correspond to the summed nucleon numbers above the respective shell gap numbers (Z = 28, N = 50) when both the J$_{pfg}$(21/2) and the J$_{dghs}$(31/2) pseudo-shells are half-filled. Our assumption is confirmed by the fact that there is no evidence indicating a pseudo-subshell formation below N = 50 as shown in Fig. 3. We summarize the results concerning the proposed pseudo-subshells and their characteristics in Table 1.

It is instructive to discuss the phase transitions for Z = 36, Kr as focusing on the critical point of Z = 36, N = 60. The reason is that the Z = 36, just close to Z = 38, would be likely to occur a phase transition at N = 60 like that at Z = 38. Interestingly, in the case of the systematic trend between Z = 36 and 38, as shown in Fig. 1(c), the point at Z = 36, N = 60 seems to be abnormal. On the other hand, for the case of the transition of N = 58 to 60 at the same Z = 36, as shown in Fig. 1(b), is likely normal. If the E(2$^+$) value at Z = 36, N = 60 is correct, there would be no phase transition in the Kr nuclei. From this, we can draw an important conclusion that *shape phase transitions at N = 60, leading to the ferro-deformation, should be given only at the Z = 38 and 40*.

Table 1. Summary of pseudo-shell configurations between double-shell closures of 28 ≤ Z ≤ 50 and 50 ≤ N ≤ 82: pseudo-shells, involving subshells, total spins, half-filled nucleon numbers, and distinctive deformation configurations.

| pseudo-shells | combined subshells | pseudo-shell total spins: nucleon capacitance | half(full)-fill summed nucleon number | representative deformation configurations: (Z, N) |
|---|---|---|---|---|
| $J_{pf}(9/2)$ | $2p_{3/2}1f_{5/2}$ | 9/2: 10 | 32 (38) | (32, N) |
| $J_{pf}(11/2)$ | $2p_{3/2}1f_{5/2}2p_{1/2}$ | 11/2: 12 | 34 (40) | (34, N) |
| $J_g(9/2)$ | $1g_{9/2}$ | 9/2: 10 | 44 or 46 (50) | (44, 56), (46, 56) |
| $J_{pfg}(21/2)$ | $2p_{3/2}1f_{5/2}2p_{1/2}1g_{9/2}$ | 21/2: 22 | 38 or 40 (50) | (38, 62), (40, 64) |
| $J_{dg}(13/2)$ | $2d_{5/2}1g_{7/2}$ | 13/2: 14 | 56 (64) | (40, 56) |
| $J_{dgh}(25/2)$ | $2d_{5/2}1g_{7/2}1h_{11/2}$ | 25/2: 26 | 62 or 64 (76) | (38, 62), (40, 64) |
| $J_{dghs}(27/2)$ | $2d_{5/2}1g_{7/2}1h_{11/2}3s_{1/2}$ | 27/2: 28 | 64 (78) | (38, 64), (40, 64) |
| $J_{dghs}(31/2)$ | $2d_{5/2}1g_{7/2}1h_{11/2}3s_{1/2}2d_{3/2}$ | 31/2: 32 | 66 (82) | (38, 66), (40, 66) |

## 3. Shape phase transitions in the two pseudo-shell regimes; $J_{pfg}(21/2)$ and $J_{dghs}(31/2)$.

For understanding a universal shell structure characteristic, we make a comparison in terms of a deformed parameter, the R values, between two proton regimes, one is from Z = 28 to 50 and the other is from Z = 50 to 82. Figures 4 and 5 summarize the R value systematic features for the two regimes.

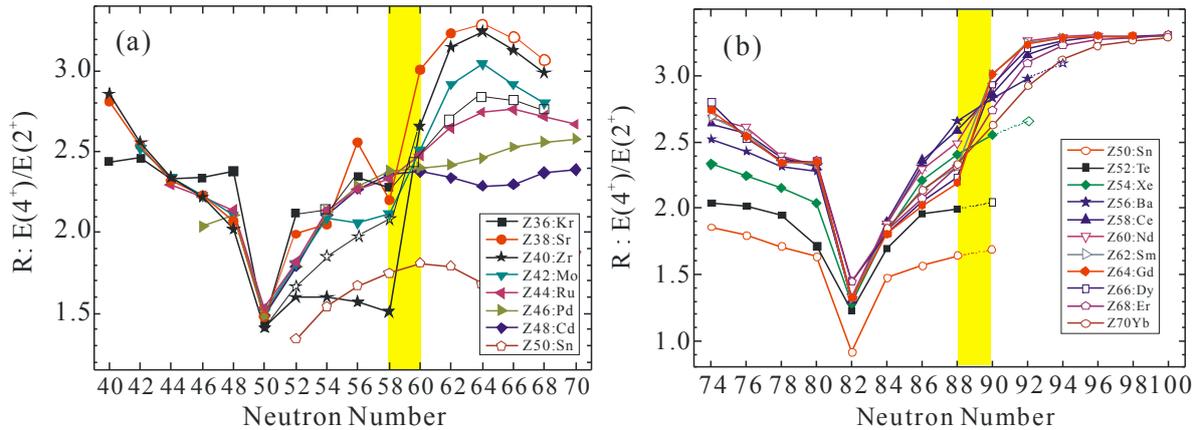

Fig. 4. Plots of the systematics for the R (= $E(4^+)/E(2^+)$) values as a function of neutron numbers in isotopes between (a) Z = 36 to 50 and (b) Z = 50 to 70 [9]. The shaded regions include the respective critical points at which shape phase transitions are occurred. The points connected with dotted lines are an expected value following the systematics while the open stars at N = 52, 54, 56, and 58 for Z = 40 denote an assumed smooth variation following the case of Z = 64.

We find a similar pattern related to shape phase transitions between the two proton regimes; $J_{pfg}(21/2)$ and $J_{dghs}(31/2)$. As evident in Figs. 4(a) and (b), we find close relations such that the N = 58 and 60, on one hand, correspond to the N = 88 and 90, and on the other hand the Z = 38 to 42 correspond to the Z = 62 to 66. From Fig. 5, we also readily find such a similar correlation between Z = 38, 40 and Z = 64, 66, as shown by a shaded region in Fig. 5(a) and in Fig. 5(b), respectively. Even the similarity between them is in general, the characteristic at Z = 40 is more dramatic and attracts more our attention than at Z = 64. A drop-valley at Z = 40 is sustained only in the range from N = 50 to 58 while it is not formed below N = 50. Then at N = 60, the valley abruptly disappears. This is a representative example of shape phase transitions over the nuclear mass chart. If we make a path following a smooth variation like the Z = 64, as shown in Fig. 4(a), such a dramatic transitional pattern would not appear. According to this assumption, a typical vibrational structure would appear to be in an excited state of Zr, with N = 56 and 58, respectively. This is the case, indicating a possibility of triple shape coexistence in these particular nuclides. In a sense, *the nucleon number 40 is a real magic number*.

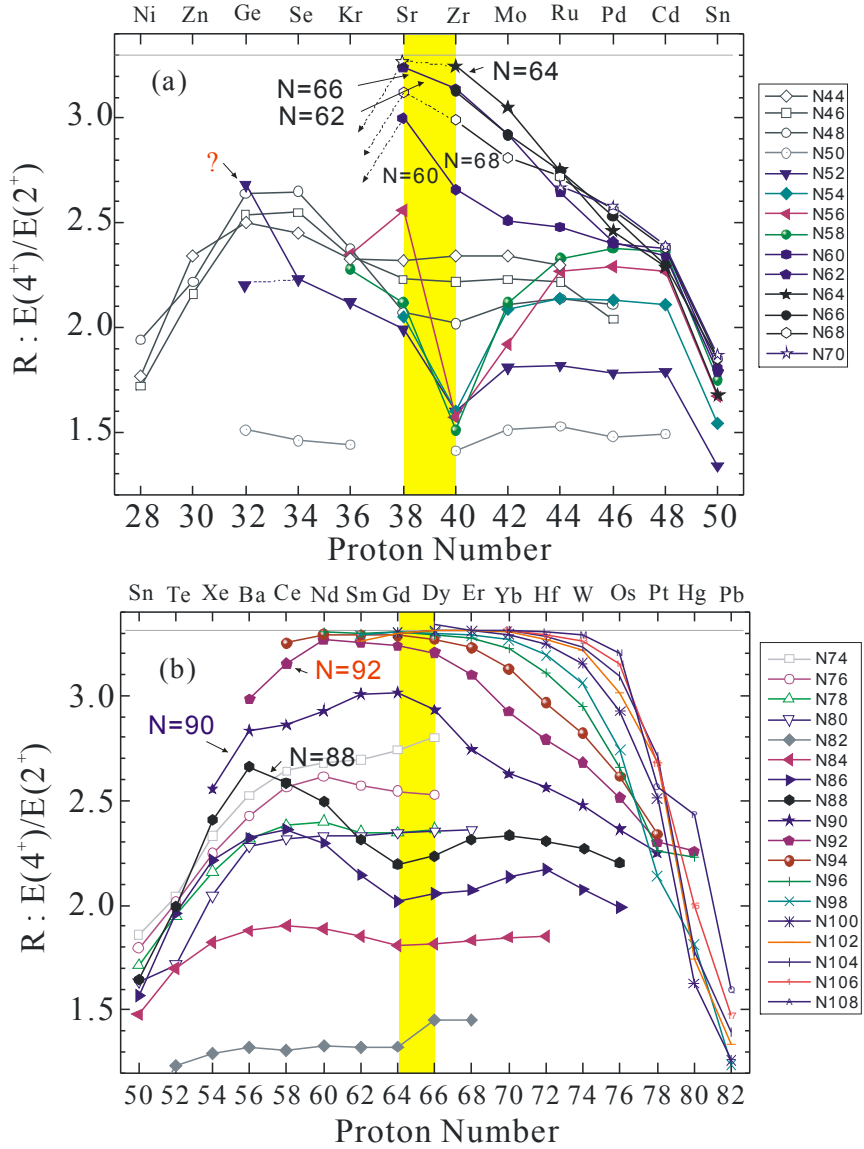

Fig. 5. Systematic plots for R (= $E(4^+)/E(2^+)$) values as a function of proton numbers in isotones between (a) N = 44 and 70, and (b) N = 74 and 108 [9]. The shaded regions indicate the critical points occurring shape phase transitions. The points with dotted lines are the expected values according to the systematic trends. It should be noted that the point at Z = 32 and N = 52, denoting a question mark, shows an anomaly, indicating a false data at the current NNDC.

As already discussed earlier, the strong isospin dependent interactions due to the proton-neutron spin-orbital doublet; protons in the lower lying (higher spin part) orbital, and neutrons in the higher lying (lower spin part) orbital, play a decisive role in controlling nuclear shape phase transitions, yielding both surface effects and shell effects. This interaction especially dominates in the nuclei near the center region of the two shell regimes; one is 28 < Z < 50, 50 < N < 82, and the other is 50 < Z < 82, 82 < N < 126. There are particularly favored combinations that have a ferro-defromation; Z = 38, 40, N = 64 in the regime of $J_{pgh}(21/2)$ and Z = 64, 66, N = 102, 104 in the regime of $J_{dghs}(31/2)$. A broad distribution of the ferro-deformation in the latter regime is most likely due to more dense level states than those of the former one. The present comparison provides clear classification about the ferro-deformation and amplifies the importance of isospin dependent spin-orbital interactions for nuclear shape phase transitions over the nuclear chart.

In order to more visually exhibit shape phase transitions, it is also instructive to draw the R values variance between two adjacent isotopes, which is shown in Fig. 6. Here the R value derivative, ΔR(N+2, N) corresponds to the R values difference between two adjacent isotopes. This parameter gives an information about the intensity for shape phase transitions. Of most noticeable is a sharp peak at Z = 40 for ΔR(60, 58). This striking increase at Z = 40 indicates a clear shape phase transition when N = 58 to 60. Furthermore, we can see a phase transition at Z = 38. As a matter of fact, if we make an assumed line with the peak at Z = 39, not shown in this even Z number, the curve for R(58, 60) reveals a symmetrical shape like that for ΔR(90, 88) in the 50 < Z < 82 domain [9]. Interestingly, it looks like a resonant functional encountered frequently in the physics world. We notice, moreover, for the case of Z = 38, there are pronounced fluctuations such as; +0.5 for ΔR(56, 54), −0.5 for ΔR(58, 56), +0.9 for ΔR(60, 58), and +0.2 for ΔR(62, 60).

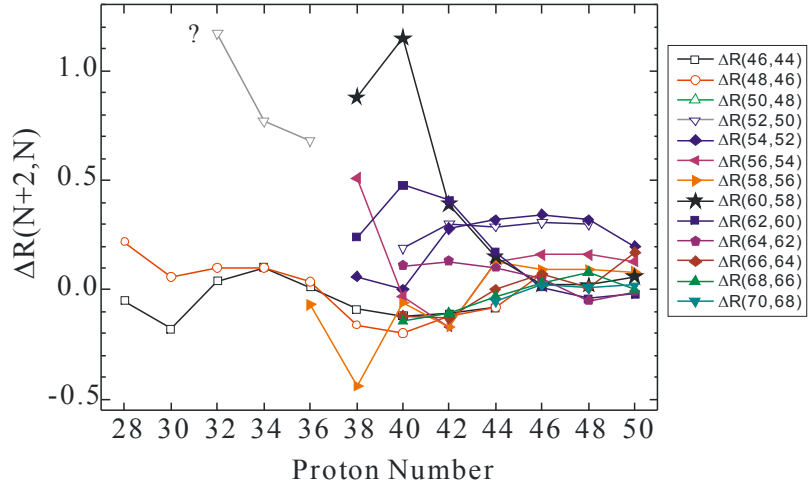

Fig. 6. The R values difference, ΔR(N+2, N) between two adjacent isotopes with Z ranging from 28 to 50. It should be noted that the peak, denoting a question mark, at Z = 32 in ΔR(52, 50) is suggested to be caused by a false data at the current NNDC library [7]. See also Fig. 5(a). Notice that the R(50, 48) values are below −0.5.

Based on the above characteristics, we leave the following remarks: First, considering such a big difference of R values, shape coexistence is expected to be in Sr and Zr with both N = 58 and 60. This is the same result as that obtained earlier. Furthermore, in view of such fluctuations at Z = 38 in the ΔR(58, 56) and ΔR(56, 54), another shape coexistence is suggested to be in Sr with N = 56. The associated feature for the number 56 is also seen at Z = 56 in Fig. 5(b), which reflects a semi-shell gap effect. Moreover, we suggest for Z = 42, Mo with N = 60 and 62 that a triple shape coexistence may be possible such that a vibrator, a triaxial rotator, and a strong rotator, giving further insights into a universal shape coexistence in the nuclei.

## 4. Conclusions

With the experimental observables related to the ground-state energy levels, we investigated the distinctive features in terms of the first $2^+$ excited energies, $E(2^+)$ and their energy ratios to the first $4^+$ levels, $R = E(4^+)/E(2^+)$, for the even-even nuclei, over $28 \leq Z \leq 50$ for protons, and $40 \leq N \leq 82$ for neutrons. By introducing the so-called *pseudo-shell* configurations built on the combined subshells, we explained various nuclear structure properties; a semi-double shell closure, a shape phase transition, and a reinforced deformation. The reinforced deformation arises when Z = 38 or 40 correlates with N = 60 and reaches R = 3.3, over Z = 38 to 42, N = 62 to 66. We defined this reinforced deformation 'a **ferro-deformation**' like a ferro-magnetism in condensed matter physics. This result implies that all the subshells between $28 \leq Z \leq 50$ and $50 \leq N \leq 82$ are involved coherently to build a big pseudo-shell with the capacitance of J = 21/2 and J = 31/2, respectively, giving rise to the ferro-deformation under half-fill conditions. We suggested that shape coexistence would occur, such as a ferro-deformation, with a strong rotational mode, and a near spherical shape, with a vibrational mode, in the transitional regions at N = 58, 60, and 60 for the nuclei, with Z = 38 or 40; $^{96}$Sr, $^{98}$Sr, $^{100}$Sr, and

$^{98}$Zr, $^{100}$Zr, and $^{102}$Zr. We suggested that the ferro-deformation should be closely associated with a strong spin-orbital interaction between neutrons in the **g$_{7/2}$** orbital and protons in the **g$_{9/2}$** orbital. Such an isospin dependent spin-orbital interaction, with the same angular momentum, $l = 4$, reinforces nuclear surface toward a sudden and dramatic deformation, giving rise to the ferro-deformation at the critical point, Z = 40, N = 64. We pointed out the similarities and differences of the ferro-deformation between two critical points; Z = 40, N = 64 and Z = 64, N = 104 and provided a clear classification of the ferro-deformation in the two regimes; one is 28 < Z < 50, 50 < N < 82 and the other is 50 < Z < 82, 82 < N < 126. It is evident that this isospin dependent spin-orbital interaction should play a decisive role, through the allowed Gamow-Teller beta transitions, in the nucleosynthesis of neutron-rich nuclei on the paths of rapid neutron capture processes.

**Note added 1**: The present work has been done in the framework of phenomenological arguments with the aid of nuclear data accumulated at the national nuclear data center, NNDC. Many references, possibly related to the present work in the literature, could not be quoted because of the referred data mostly from NNDC.
**Note added 2**: Any abbreviation is avoided since it makes the readers confusing like a jargon.